\documentclass[journal]{IEEEtran}

\usepackage[T1]{fontenc}

\DeclareUnicodeCharacter{03A4}{T}

\ifCLASSINFOpdf
\else
\fi

\usepackage{amsfonts}
\usepackage{algorithmic}
\usepackage[ruled,vlined]{algorithm2e}

\usepackage{adjustbox}
\usepackage{graphicx}
\usepackage{caption}
\usepackage[justification=centering]{caption}
\usepackage[parfill]{parskip}
\usepackage{amsmath}
\usepackage{subcaption} 
\usepackage{placeins} 

\usepackage{booktabs} 
\usepackage{siunitx} 
\usepackage{makecell} 

\usepackage{booktabs}
\usepackage{adjustbox}
\usepackage{multirow}
\usepackage[caption=false,font=normalsize,labelfont=sf,textfont=sf]{subfig}
\usepackage{stfloats}
\usepackage{url}
\usepackage{verbatim}
\usepackage{cite}
\usepackage{svg}

\usepackage{booktabs}
\usepackage{multirow}
\usepackage{array}
\usepackage{tabularx}
\usepackage{color,soul}

\usepackage{subcaption}

\begin{document}

\title{Deep Reinforcement Learning Orchestration of Game-Theoretic User Association and Resource Allocation in HetNets}

\author{\IEEEauthorblockN{
Sotiris Kopsinos,
Alexandros I. Papadopoulos,
Antonios Lalas, 
Konstantinos Votis,
and Christos Liaskos 
}
\thanks{S. Kopsinos and A. Papadopoulos are with the Computer Science Engineering Department, University of Ioannina, Ioannina, Greece and with the Information Technologies Institute, CERTH, Greece (e-mails: s.kopsinos@uoi.gr/skopsinos@iti.gr, a.papadopoulos@uoi.gr/alexpap@iti.gr).}
\thanks{A. Lalas and K. Votis are with the Information Technologies Institute, CERTH, Greece (e-mails: \{lalas, kvotis\}@iti.gr).}
\thanks{C. Liaskos is with the Computer Science Engineering Department, University of Ioannina, Ioannina, and with the Foundation for Research and Technology Hellas (FORTH), Greece (e-mail: cliaskos@uoi.gr).}
\thanks{This work was funded by the SNS JU under the EU Horizon Europe research and innovation program through the NATWORK project (Grant No. 101139285).}
}

\maketitle

\begin{abstract}
Managing dynamic User Association and Resource Allocation (UARA) in modern Heterogeneous Cellular Networks (HetNets) remains a critical open challenge. Existing mathematical optimization and Reinforcement Learning approaches face limitations in handling low-latency decision-making under dynamic traffic conditions. This paper introduces a novel orchestration scheme for game-theoretic UARA in HetNets. The proposed bilevel framework distributes UARA decisions to User Equipment through a multi-objective non-cooperative game. Overlaying the distributed game, a centralized Deep Reinforcement Learning controller orchestrates network performance by dynamically configuring the game's utility parameters, enabling transitions between power awareness, coverage enhancement, and balanced operation. Evaluated on urban HetNet topologies with 3GPP TR 38.901-compliant channel modeling, the proposed framework closely approximates the optimal policy for the considered operational objectives, while delivering higher network throughput than conventional association methods. Furthermore, it incurs low computational overhead and maintains stable performance across the evaluated traffic densities without retraining.

\end{abstract}


%
\IEEEpeerreviewmaketitle

\section{Introduction}
The rapid growth  in the number of mobile devices and traffic demand in modern cellular networks~\cite{agiwal2016next, vaezi2022cellular} has driven network deployments beyond traditional macro-only Base Station (BS) architectures~\cite{agiwal2016next, liu2016user}. This evolution has led to multi-tier BS deployments, commonly referred to as Heterogeneous Cellular Networks (HetNets)~\cite{liu2016user, xu2021survey}. By deploying small-cell BSs, such as Pico BSs (PBSs) and Femto BSs (FBSs), within existing macro-cell coverage, HetNets enable dense spectrum reuse, thereby increasing network capacity in a cost-effective and energy-aware manner\cite{liu2016user,xu2021survey}.

As key enablers of Beyond 5G (B5G) systems, HetNets integrate heterogeneous technologies to improve coverage, capacity, and reliability \cite{wang2023road}. Massive Multiple-Input Multiple-Output antennas are widely deployed, providing spatial multiplexing gains and increased capacity \cite{wang2023road}. Moreover, higher frequencies, such as millimeter wave bands, provide wide bandwidths for data-intensive applications \cite{wang2023road}. In this context, HetNets could also incorporate emerging programmable wireless technologies, such as Reconfigurable Intelligent Surfaces (RISs), which enable energy-efficient control of the propagation environment. Their integration into HetNets can provide benefits such as coverage extension, blockage mitigation, and interference management.\cite{11021423}.

Despite their importance for B5G, HetNets introduce challenging operational trade-offs due to the coexistence of high-power and low-power BSs~\cite{xu2021survey}. Conventional user association based on maximum received signal power tends to concentrate User Equipment (UEs) at macro cells, often leading to small-cell under-utilization, reduced system capacity, and increased power consumption~\cite{liu2016user,ye2013user}. On the other hand, offloading UEs to small cells can increase inter-cell interference from the macro-tier and degrade Quality-of-Service (QoS)~\cite{liu2016user, ye2013user}. Therefore, efficient HetNet operation requires jointly addressing user association and resource allocation to balance load, manage interference, reduce power consumption, and satisfy diverse UE service requirements. 

Nevertheless, several existing frameworks for joint User Association and Resource Allocation (UARA) in HetNets face practical limitations under dynamic network conditions. In particular, optimization-based methods often require iterative computations~\cite{khalili2020joint, liu2019joint, kim2017qos, soleimani, rezvani2021resource}, which may become costly for real-time operation. Reinforcement Learning (RL)-based algorithms, on the other hand, can reduce part of the computational burden through offline training, but they often rely on combinatorial action spaces~\cite{yang2022distributed, liu2024reinforcement} and user-dependent state space dimensions~\cite{yang2022distributed, liu2024reinforcement, allagiotis2023reinforcement, zhao2019deep}, and may still incur high computational complexity as network density scales. Moreover, many optimization and RL formulations depend on detailed UE-specific information, such as CSI data or complete SINR vectors~\cite{yang2022distributed, dong2023energy, ozcan2020robust, soleimani, rezvani2021resource}. Since such measurements can support accurate UE localization~\cite{10355064, cigno2022integrating, de2022positioning}, their transmission can raise privacy concerns~\cite{10355064, cigno2022integrating}. In addition, MARL schemes that model UEs as agents may require local neural network (NN) inference~\cite{zhao2019deep}, which can be unsuitable for resource-constrained devices. Another limitation is that many RL models employed in UARA schemes are designed for fixed numbers of UEs~\cite{zhao2019deep, liu2024reinforcement, yang2022distributed}, complicating their adaptation to varying traffic densities. Furthermore, several works assume uniform UE distributions~\cite{chinipardaz2025user, zhao2019deep, allagiotis2023reinforcement, trabelsi2017user}, although practical HetNets often feature zones of concentrated traffic. Additionally, evaluations based on conventional statistical fading models~\cite{yang2022distributed, chen2025maddpg, 10264114} may not fully capture spatially consistent large- and small-scale propagation effects. Finally, many UARA frameworks optimize fixed objectives~\cite{chen2025maddpg, trabelsi2017user, 9187838, 11096913, 10264114}, while practical networks may need to adapt their operational goals depending on congestion levels.

Motivated by the aforementioned limitations, we propose a novel orchestration scheme for game-theoretic UARA in HetNets. By overlaying a centralized Deep Reinforcement Learning (DRL) controller on a decentralized game-theoretic execution layer, the proposed framework provides scalable multi-objective resource management. The main contributions of our work are summarized as follows:
\begin{itemize}
\item We introduce a multi-objective non-cooperative game formulation for HetNets, where UARA decisions are distributed to UEs. The proposed utility function jointly accounts for user data rate, SINR, and BS power consumption. The resulting UE-side computations require simple algebraic operations, making the framework suitable for a broad range of devices.

\item We propose an orchestration scheme to control the performance of the distributed game. In particular, we employ a low-complexity centralized DQN controller, which configures the utility parameters determining the relative importance of rate, signal quality, and power cost across UEs. In this way, the controller can dynamically steer the network toward different macroscopic operational objectives. The proposed DQN relies on high-level BS load distribution data as inputs, thereby reducing the need for UE-specific information at the controller side.

\item We evaluate the proposed DQN-orchestrated framework in a practical HetNet scenario under 3GPP-standardized urban channel conditions. Simulation results show that it supports multi-objective resource management, provides high network throughputs, and operates with low computational overhead, while maintaining good performance across traffic loads not encountered during training.
\end{itemize}

The remainder of this paper is organized as follows. Section~\ref{sec:lit} provides an overview of the literature. Section~\ref{sec:system_model} describes the network and communication model. Section~\ref{sec:game_form} formulates the distributed, multi-objective UARA game. Section~\ref{sec:metacontrol} presents the proposed DQN-based scheme, which orchestrates the UARA game. Section~\ref{sec:per_eval} evaluates the performance of the proposed framework in a practical HetNet environment. Finally, Section~\ref{sec:conclusions} concludes the paper.

\section{Literature Overview}\label{sec:lit}
Mathematical optimization has been widely used to address the joint UARA problem. One main direction focuses on maximizing system throughput~\cite{soleimani, khalili2020joint, rezvani2021resource}, while several works study energy-efficient HetNet operation under user QoS constraints~\cite{kim2017qos, ni2024user, dong2023energy}. Since the joint UARA problem is typically non-convex and combinatorial, existing optimization-based methods commonly rely on reformulation techniques. These include relaxing binary association variables into continuous ones~\cite{kim2017qos, dong2023energy, liu2019joint} and decomposing the original problem into more tractable subproblems~\cite{khalili2020joint, liu2019joint, kim2017qos,ni2024user, dong2023energy}. To reduce the computational overhead of joint online optimization, the authors in~\cite{ozcan2020robust} separated the two processes, using mathematical optimization solely for offline resource allocation, while relying on low-complexity heuristics for online user association and scheduling.

In parallel, RL-based methods have been proposed to shift part of the UARA computational burden to offline training. The first RL-based framework for joint UARA in HetNets was introduced in~\cite{zhao2019deep}. In this work, UEs are modeled as independent agents in a multi-agent RL (MARL) framework, which maximizes long-term aggregate network utility while satisfying QoS constraints. MARL was also considered in~\cite{yang2022distributed}, where a multi-agent dueling deep Q-network with distributed coordinated learning was used for joint device association, spectrum allocation, and power allocation. Unlike UE-centric approaches, this framework models BSs as learning agents and aims to maximize the network data rate under QoS constraints. In~\cite{chen2025maddpg}, the authors proposed a MARL framework for UAV-assisted HetNets. After an initial matching game and a Lagrangian-dual-based user association step, UAVs and terrestrial BSs act as autonomous agents that dynamically allocate slicing resources to balance throughput, latency, and coverage. Centralized RL schemes have also been employ for joint UARA. The authors in~\cite{allagiotis2023reinforcement} proposed a centralized deep Q-network (DQN) controller that jointly assigns users to BSs and allocates Physical Resource Blocks. Following this centralized paradigm, the framework in\cite{liu2024reinforcement} addresses end-to-end network slicing in HetNets. While Dijkstra’s algorithm handles core network link assignment, centralized DQN agents iteratively optimize user association, wireless bandwidth allocation, and Virtual Network Function placement under user QoS constraints. 

Game Theory is another strong candidate for UARA, as it decomposes the global problem into local player interactions. In~\cite{trabelsi2017user}, the authors proposed a non-cooperative exact potential game, with individual BSs as players. This centrally coordinated framework uses iterative best-response dynamics to optimize Cell Individual Offsets for user association alongside power transmission patterns. Other works consider both UEs and BSs as players. For instance,~\cite{10264114} proposes a hierarchical framework that combines an evolutionary game for user association with a Stackelberg differential game for resource allocation in wireless-backhauled HetNets. At the user level, UEs dynamically select BSs via the evolutionary game. At the resource level, the MBS determines backhaul prices as the leader, while small cells adapt their bandwidth and access-throughput strategies as followers. 
Several works adopt UE-centric non-cooperative games. For example,~\cite{9187838} models spectrum allocation and user association as coordinated non-cooperative exact potential games. Regarding user association, UEs optimize their association time through greedy best response dynamics, balancing spectral efficiency and load. Similarly,~\cite{11096913} models joint user association and channel selection as a non-cooperative stochastic game among UEs. To address the non-convexity of the problem under imperfect Channel State Information (CSI), the authors proposed a multi-agent double deep Q-network algorithm, enabling UEs to maximize aggregate network rate subject to QoS constraints.

\section{Network and Communication Model} \label{sec:system_model}
\begin{figure}[t]
 \centering  \includegraphics[width=\linewidth]{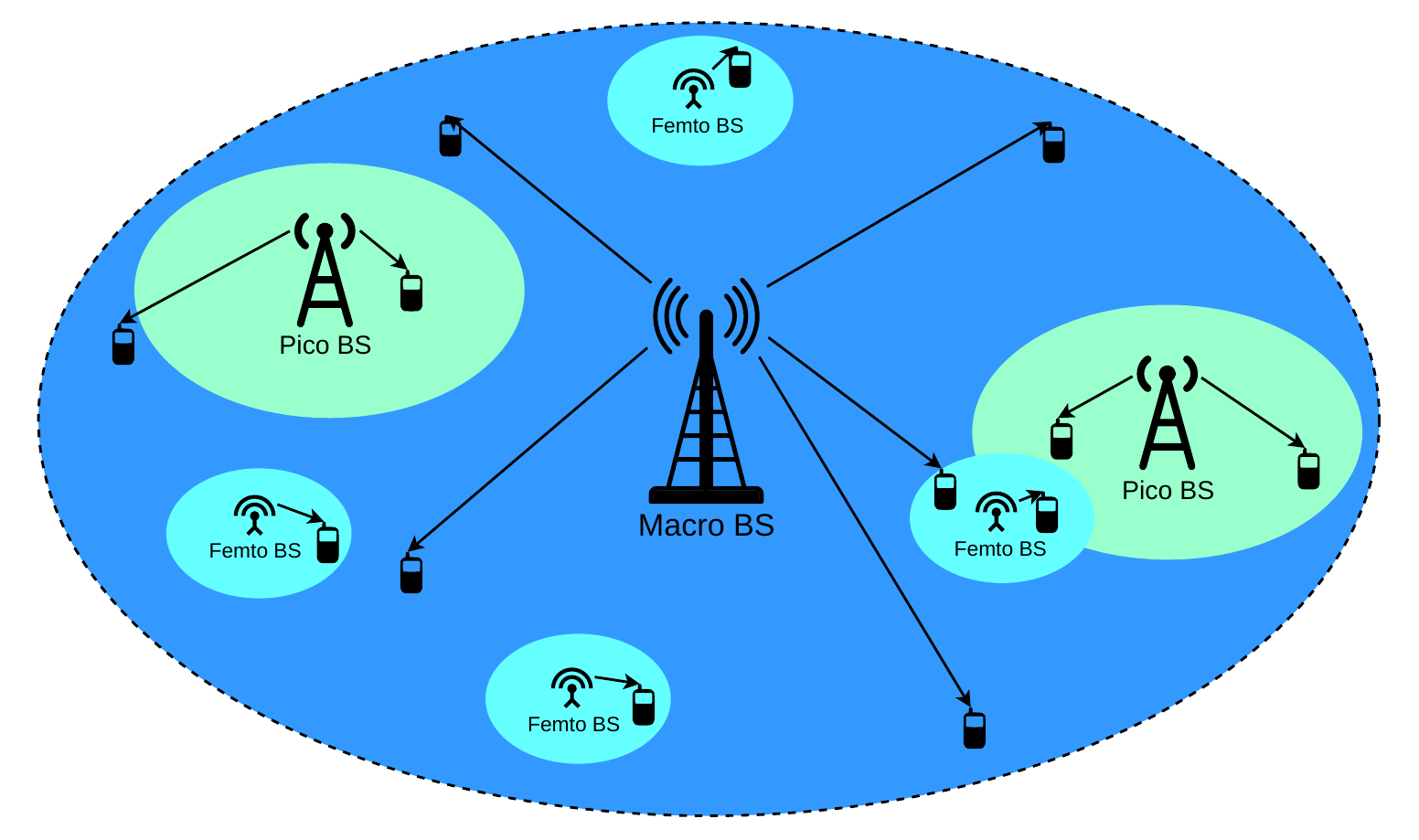}
   \caption{A three-tier heterogeneous cellular network.}
\label{fig:hetnet}
\end{figure}

We consider a three-tier HetNet consisting of one MBS, $P$ PBSs, and $F$ FBSs, as shown in Fig.~\ref{fig:hetnet}. Each BS operates on $K$ shared orthogonal channels of bandwidth $W$. A total of $N$ users are randomly distributed across the network area. The sets of UEs and available channels are denoted by $\mathcal{U} = \{1, 2, \dots, N\}$ and $\mathcal{K} = \{1, 2, \dots, K\}$, respectively. The BS set is denoted by $\mathcal{B} = \{0, 1, \dots, P+F\}$, where index $0$ corresponds to the MBS, indices $1$ through $P$ to the PBSs, and indices $P+1$ through $P+F$ to the FBSs.

It is assumed that each user $u \in \mathcal{U}$ associates with exactly one BS-channel pair $(b, k) \in \mathcal{B} \times \mathcal{K}$ at any given time. To formulate this connection, we define a binary association variable $x_{u,b}^k \in \{0, 1\}$, where $x_{u,b}^k = 1$ if user $u$ connects to BS $b$ on channel $k$, and $x_{u,b}^k = 0$ otherwise. Thus:
\begin{equation}
\sum_{b \in \mathcal{B}} \sum_{k \in \mathcal{K}} x_{u,b}^k = 1, \quad \forall u \in \mathcal{U}.
\end{equation}
When user $u$ is served by BS $b$ on channel $k$, i.e., $x_{u,b}^k = 1$, the received Signal-to-Interference-plus-Noise Ratio (SINR) is given by:
\begin{equation} \label{eq:sinr}
\gamma_{u,b}^k = \frac{p_{b}^k g_{u,b}^k}{\sum_{b' \in \mathcal{B} \setminus \{b\}} p_{b'}^k g_{u,b'}^k + W N_0},
\end{equation} 
where $p_{b}^k$ is the transmission power of BS $b$ on channel $k$, and $g_{u,b}^k$ denotes the channel gain between BS $b$ and user $u$ over channel $k$. The channel gain accounts for path loss, shadowing, and fast fading effects. The term $\sum_{b' \in \mathcal{B} \setminus \{b\}} p_{b'}^k g_{u,b'}^k$ denotes the co-channel interference from all other base stations transmitting on the same channel $k$, and $N_0$ is the noise power spectral density.

Users in cellular networks typically belong to different priority tiers based on their technical QoS requirements and the importance of their service\cite{chinipardaz2025user,7533456}. Therefore, to implement unequal user priority, each user $u \in \mathcal{U}$ is assigned a priority weight $\omega_u > 0$. We define $\mathcal{U}_b^k = \{u' \in \mathcal{U} \mid x_{u',b}^k = 1\}$ as the set of all users simultaneously associated with BS $b$ on channel $k$ and $\Omega_b^k = \sum_{u' \in \mathcal{U}_b^k} \omega_{u'}$ as their aggregate priority weight. The fraction of channel resources allocated to user $u$ is strictly proportional to its priority weight relative to the aggregate weight of users sharing that specific BS-channel pair. Thus, the achievable downlink data rate for user $u$ from base station $b$ over channel $k$ is expressed as:
\begin{equation} \label{eq:rate}
R_{u,b}^k = \frac{\omega_u}{\Omega_b^k} r_{u,b}^k,
\end{equation}
where 
\begin{equation} \label{eq:unc_rate}
r_{u,b}^k=W \log_2(1 + \gamma_{u,b}^k)
\end{equation}
denotes the uncongested data rate, i.e., the theoretical rate achieved under exclusive channel access. 

\section{Game Formulation for Distributed User Association and Resource Allocation}\label{sec:game_form}
To overcome the complexity of global UARA optimization, we employ distributed decision-making. Specifically, we model UARA as a multi-objective non-cooperative game, defined by the tuple $\mathcal{G} = \langle \mathcal{U}, \mathcal{S}, \{J_u\}_{u \in \mathcal{U}} \rangle$, where:
\begin{itemize}
    \item $\mathcal{U}$ is the finite set of players, i.e., the UEs in the network.
    \item $\mathcal{S} = \prod_{u \in \mathcal{U}} \mathcal{S}_u$ is the global strategy space, where $\mathcal{S}_u$ denotes the set of available strategies for user $u$.  A strategy $s_u \in \mathcal{S}_u$ is a singleton selection corresponding to exactly one BS-channel pair, such that $s_u = (b, k)$.
    \item $J_u : \mathcal{S} \rightarrow \mathbb{R}$ is the utility function for user $u$, which maps a chosen strategy profile to a real-valued payoff.
\end{itemize}

Let $s = (s_u, s_{-u}) \in \mathcal{S}$ denote the strategy profile of all users, where $s_u$ is the strategy chosen by user $u$ and $s_{-u}$ is the strategy profile of all other users in the network. The utility function evaluates the payoff of a potential connection based on a parameterized sum of normalized communication metrics. Specifically, for a user $u$ selecting strategy $s_u = (b, k)$, the resulting utility is given by:
\begin{equation} \label{eq:util}
J_u(s) = w_r \bar{R}_{u,b}^k(s) + w_\gamma \bar{\Gamma}_{u,b}^k(s_u) - w_p \bar{P}_b(s_u).
\end{equation}
In Eq. \eqref{eq:util}, $\bar{R}_{u,b}^k \in [0, 1]$ is the achievable data rate of user $u$, normalized by the maximum uncongested data rate available in the network, and depends on the channel load induced by the complete strategy profile $s$. The term $\bar{\Gamma}_{u,b}^k \in [0,1]$, with $\Gamma_{u,b}^k = 10\log_{10}(\gamma_{u,b}^k)$, captures the received SINR in dB, min-max normalized over the network-wide SINR range. This metric reflects the radio-link quality of the selected BS-channel pair and, in contrast to $\bar{R}_{u,b}^k$, is independent of channel congestion. The term $\bar{P}_b \in[0, 1]$ represents a fixed, tier-dependent power cost associated with BS $b$, obtained by normalizing its logarithmic transmit power against the maximum BS power in the network. The weighting parameters $(w_{r}, w_{\gamma}, w_{p}) \in [0, 1]^3$ control the relative importance of data rate, signal quality, and power consumption. Including both $\bar{R}_{u,b}^k$ and $\bar{\Gamma}_{u,b}^k$ allows the utility to distinguish between connections that offer high data rates due to low congestion and connections that provide genuinely strong radio conditions. This prevents users from selecting lightly loaded but signal-poor channels, while still enabling load-aware association. Hence, the proposed utility function formulation can support diverse QoS requirements in HetNets, ranging from throughput-oriented to reliability-sensitive services.

The objective of each user $u$ is to maximize its individual utility given the strategies of other users:
\begin{equation}
\max_{s_u \in \mathcal{S}_u} J_u(s), \quad \forall u \in \mathcal{U}.
\end{equation}
The non-cooperative game reaches a stable state when no user can unilaterally improve its utility by changing its current association. Mathematically, this state corresponds to a Pure Strategy Nash Equilibrium (PSNE). A strategy profile $s^* = (s_1^*, \dots, s_N^*)$ constitutes a PSNE if and only if it satisfies the following condition:
\begin{equation} \label{eq:psne}
J_u(s_u^*, s_{-u}^*) \ge J_u(s'_u, s_{-u}^*), \quad \forall s'_u \in \mathcal{S}_u, \forall u \in \mathcal{U}.
\end{equation} 

Assuming that the transmit powers of the base stations are fixed during the UARA game, $\bar{\Gamma}_{u,b}^k$ and $\bar{P}_b$ are static metrics and independent of the strategies of other users. Thus, the coupling between players occurs strictly through the load-aware rate calculation $\bar{R}_{u,b}^k$, where the resource fraction allocated to each user depends on the user's priority weight relative to the total weight of all users sharing the same BS-channel pair. 
This particular dependence maps $\mathcal{G}$ to the class of weighted congestion games with player-specific constants. For the considered weighted congestion structure, the existence of a PSNE is not guaranteed in full generality. To establish strict mathematical  convergence within a finite number of steps, we shift our objective to an $\epsilon$-Approximate PSNE ($\epsilon$-PSNE),  defined as a state in which no user can improve its utility by more than a dynamic hysteresis threshold $\tau$.

\textbf{\textit{Theorem 1:}} Consider the UARA game $\mathcal{G}$. Defining the strategy update rule so that a user only switches its association if its individual utility gain satisfies $\Delta J_u > \tau$, where $\tau = w_r \omega_u / (\Omega_n + \omega_u)$, maps the formulation to a Generalized Ordinal Potential Game. This guarantees monotonic convergence to an $\epsilon$-PSNE in a finite number of steps.

\textbf{\textit{Proof:}} Let the global potential function $\Phi(s)$ be the sum of all individual user utilities:
\begin{equation}
\Phi(s) = \sum_{v \in \mathcal{U}} J_v(s).
\end{equation}
Assume user $u$ unilaterally switches from an old association $o$ to a new one $n$, which corresponds to a specific BS-channel pair $(b, k)$. We denote $\Omega_n = \Omega_b^k$ as the aggregate priority weight at the target channel immediately prior to the association of user $u$. The network potential changes by $\Delta \Phi = \Delta J_u + G_o - D_n$, where $\Delta J_u$ is the total utility gained by user $u$, $G_o$ is the utility gained by the users left behind at $o$ (which is strictly positive as the aggregate priority weight decreases), and $D_n$ is the congestion penalty introduced to the already connected users at the target channel, denoted as $\mathcal{U}_n$.

\begin{algorithm}[t]
\caption{Distributed Game-Theoretic UARA via ABRD}
\label{alg:uara}
\textbf{Input:} $\mathcal{U}, \mathcal{B}, \mathcal{K}, \{\omega_u\}_{u \in \mathcal{U}}, \{\bar{P}_b\}_{b \in \mathcal{B}}, W, (w_r, w_\gamma, w_p)$
\textbf{Output:} Final association profile $\{x_{u,b}^k\}_{\forall u \in \mathcal{U}}$
\vspace{1mm}  
\hrule 
\vspace{1mm}
\textbf{Initialization:} \\
Measure local SINR $\gamma_{u,b}^k, \forall (b,k) \in \mathcal{B} \times \mathcal{K}$ at each UE \\
Transmit bounds $\gamma_{\min}, \gamma_{\max}, R_{\max}$ from controller \\
Initialize each UE strategy via max-SINR: $s_u = (b,k) \leftarrow \arg\max\limits_{(b',k') \in \mathcal{B} \times \mathcal{K}} \gamma_{u,b'}^{k'}$ \\
Initialize association indicators $x_{u,b}^k$  $\forall u \in \mathcal{U}$ \\
Initialize network loads $\mathcal{U}_b^k$ $\forall (b,k) \in \mathcal{B} \times \mathcal{K}$

\Repeat{$\forall u \in \mathcal{U}, J_u(s_u, s_{-u}) \geq J_u(s'_u, s_{-u}), \forall s'_u \in \mathcal{S}_u$}{
    \For{each user $u \in \mathcal{U}$ asynchronously}{
        Given $s_u = (b,k)$ \\
        Evaluate utility $J_u(s'_u, s_{-u}), \forall s'_u \in \mathcal{S}_u$ \\
        Compute best response: $s^{BR}_u = (b^{BR},k^{BR}) \leftarrow \arg\max\limits_{s'_u \in \mathcal{S}_u} J_u(s'_u, s_{-u})$ \\
        
        \If{$J_u(s^{BR}_u, s_{-u}) > J_u(s_u, s_{-u})$}{
            $s_u \leftarrow s^{BR}_u$ \\
            $x_{u,b}^k \leftarrow 0, \quad x_{u,b^{BR}}^{k^{BR}} \leftarrow 1$ \\
            Update loads: $\mathcal{U}_b^k \leftarrow \mathcal{U}_b^k \setminus \{u\}, \quad \mathcal{U}_{b^{BR}}^{k^{BR}} \leftarrow \mathcal{U}_{b^{BR}}^{k^{BR}} \cup \{u\}$
        }
    }
}
\end{algorithm}

Because the aggregate priority weight of the target connection increases from $\Omega_n$ to $\Omega_n + \omega_u$, and the individual user SINR and power cost metrics are independent of congestion, the utility reduction experienced by the existing users is caused solely by the load-aware rate component:
\begin{equation} \label{eq:Dn_penalty}
D_n = w_r \left( \frac{\omega_u}{\Omega_n (\Omega_n + \omega_u)} \right) \sum_{v \in \mathcal{U}_n} \omega_v \bar{r}_{v,n}.
\end{equation}
Given that the normalized uncongested rate satisfies $\bar{r}_{v,n} \le 1$, the sum of the weighted uncongested rates of the existing users is strictly upper-bounded by their aggregate priority weight, $\sum_{v \in \mathcal{U}_n} \omega_v \bar{r}_{v,n} \le \Omega_n$. Substituting this bound in Eq. \eqref{eq:Dn_penalty} provides the maximum possible utility reduction of the users currently connected to the target channel:
\begin{equation}
D_n \le w_r \left( \frac{\omega_u}{\Omega_n + \omega_u} \right).
\end{equation}
By defining this upper bound as our dynamic threshold $\tau$ and applying the condition $\Delta J_u > \tau$ to our game, it is guaranteed that $\Delta J_u > D_n$, which ensures $\Delta \Phi > 0$. Since the strategy space is finite and every update strictly increases the global potential, the game must terminate at an $\epsilon$-PSNE in a finite number of steps. \hfill \IEEEQEDopen

Although condition $\Delta J_u > \tau$ guarantees theoretical convergence, it leads to overly conservative player strategies. In practice, unconstrained best-response dynamics for games of this type have been shown to consistently converge to an equilibrium\cite{caballero2017multi}. Therefore, to bridge the theoretical worst-case bound with practical execution, we adopt unconstrained Asynchronous Best-Response Dynamics (ABRD), where users update their strategies whenever $\Delta J_u > 0$. 

\begin{figure*}[t]
 \centering
 \includegraphics[width=0.87\textwidth, trim=28 0 0 0, clip]{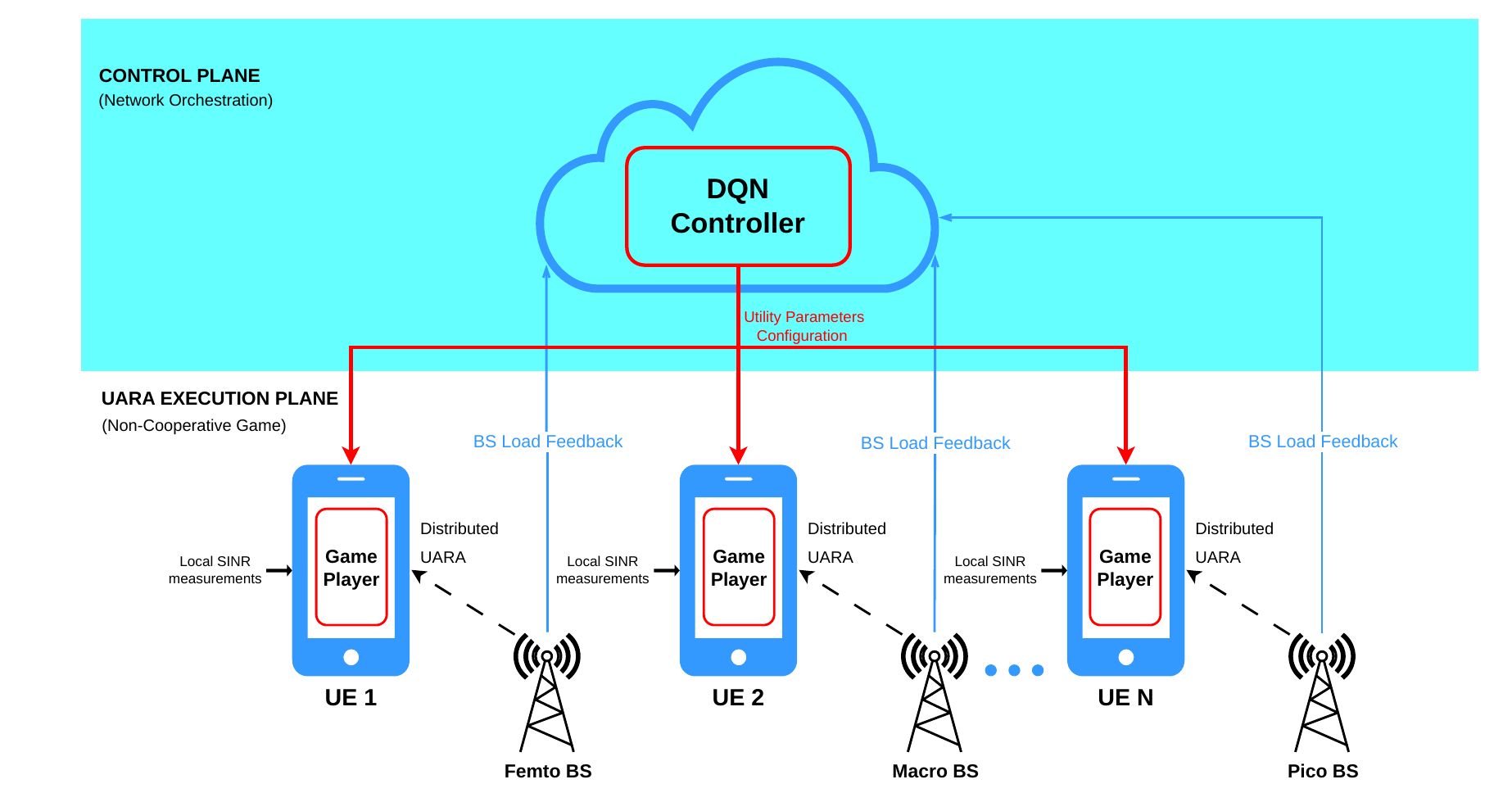}
 \caption{Illustration of the Proposed DQN-based Orchestration Scheme for Game-Theoretic UARA.}
 \label{fig:arch}
\end{figure*}

The practical execution of the distributed UARA game using ABRD is presented in Alg.~\ref{alg:uara}. Initially, UEs measure their local SINR for all available BS-channel pairs. Each UE then determines its minimum and maximum SINR and transmits only these two values to the network controller—without associating them with specific links. The controller then evaluates the global extreme values across the network to define the bounds $\gamma_{\min}$, $\gamma_{\max}$, and $R_{\max}$, which it broadcasts back to the UEs so that they normalize their individual data rate and SINR metrics. Here, $R_{\max}$ denotes the maximum possible uncongested data rate, calculated using $\gamma_{\max}$ in Eq. \eqref{eq:unc_rate}. Following, the UEs establish their initial connections based on the Max-SINR association rule. They then begin the distributed UARA game. During the game, each UE asynchronously evaluates its utility across its strategy space. If a strictly positive utility gain is identified, the UE unilaterally updates its association to the corresponding BS-channel pair, and the global channel loads are updated accordingly. This asynchronous process continues until no UE can further improve its utility by changing its current connection, at which point the network converges to a PSNE. Because each UE evaluates all $|\mathcal{B}| \times |\mathcal{K}|$ links, the per-iteration computational complexity of Alg.~\ref{alg:uara} is $\mathcal{O}(N |\mathcal{B}| |\mathcal{K}|)$, scaling linearly with each network dimension.

\section{Dynamic Parameter Tuning for Game-Theoretic UARA with Reinforcement Learning} \label{sec:metacontrol}
As shown in Eq. \eqref{eq:util}, the resulting strategy profile of the UEs and, consequently, also the overall network performance are determined by the weighting parameters $(w_{r}, w_{\gamma}, w_{p}) \in [0, 1]^3$ of the game's utility function. Each combination of these values balances the trade-offs between throughput, signal quality, and power consumption in a distinct way, both at the microscopic and macroscopic level.

However, analytically computing the optimal weights for a desired macroscopic network objective (e.g., prioritizing network throughput or coverage, or reducing power consumption) is far from straightforward. To the best of our knowledge, no closed-form expression can directly map desired aggregate network performance metrics to exact utility parameter values, since user associations result from a non-cooperative game, which also does not admit a guaranteed unique equilibrium. Furthermore, relying on computational methodologies such as exhaustive search would require simulating the game to convergence for every candidate parameter configuration, hindering low-latency network orchestration. 
Consequently, a model-free approach is required to dynamically map macroscopic network objectives to optimal utility parameters without excessive computational overhead. To this end, we propose a network orchestration scheme based on Reinforcement Learning, which tunes the weighting parameters of the multi-objective utility function.

\subsection{MDP Formulation for Network Orchestration}
\label{sec:mdp}
To address the dynamic weighting parameter configuration problem, without the constraints of tabular Q-learning in complex action and state spaces, we employ Deep Reinforcement Learning, where a centralized network controller acts as the agent. The orchestration process is modeled as a Markov Decision Process (MDP) defined by the tuple $\langle \mathcal{X}, \mathcal{A}, \mathcal{P}, \mathcal{R}, \delta \rangle$:

\begin{itemize}
    \item State Space ($\mathcal{X}$): The DRL agent receives real-time feedback on the load distribution across the network. We define the state $x_{(t)} \in \mathcal{X}$ at time step $t$ as a vector consisting of the standardized load fraction of every BS and the weighting parameters previously applied:
    \begin{equation}
    x_{(t)}=
    \left[\{\bar{L}_{b(t)}\}_{b\in\mathcal{B}},
    w_{\gamma(t-1)},w_{p(t-1)}\right].
    \end{equation}
    The raw load fraction $L_{b(t)}$ of each BS $b \in \mathcal{B}$ is calculated as the ratio of its associated users to the total number of users $N$:
    \begin{equation}
    L_{b(t)} = \frac{\sum_{k \in \mathcal{K}} |\mathcal{U}_{b(t)}^{k}|}{N}.
    \end{equation}
    To ensure gradient stability, the raw load fraction is standardized using a tier-specific mean $\mu_b$ and standard deviation $\sigma$:
    \begin{equation} \label{eq:stand}
    \bar{L}_{b(t)} = \frac{L_{b(t)} - \mu_b}{\sigma}.
    \end{equation}
    This step ensures consistent scaling across base station tiers, as directly feeding heterogeneous fractions into the state space would bias the learning process and lead to unstable training.
    
    \item Action Space ($\mathcal{A}$): In our framework, we fix $w_{r}$ and tune $(w_{\gamma}, w_{p}) \in [0, 1]^2$. This design choice allows the controller to guide the network toward different objectives, such as power-aware and coverage-enhancing operation. Moreover, it reduces the dimensionality of the optimization space, resulting in a less complex problem and facilitating low-latency policy implementation, which is a priority for the design of our proposed scheme. Therefore, at each time step $t$, the DRL agent selects an action $a_{(t)}$, which represents a configuration of $(w_{\gamma}, w_{p}) \in [0, 1]^2$. To avoid the computational overhead of continuous optimization, the utility parameters are discretized into the following finite set:
    \begin{equation} \label{eq:step}
    \mathcal{W}_\gamma = \mathcal{W}_p = \{0, \Delta, 2\Delta, \dots, 1\}. 
    \end{equation}
    Thus, the two-dimensional discrete Action Space is defined as $\mathcal{A} = \mathcal{W}_\gamma \times \mathcal{W}_p$, and the action vector as:
   \begin{equation}
    a_{(t)} = \left[ w_{\gamma (t)}, w_{p (t)} \right] \in \mathcal{A}.
    \end{equation}
    
    \item Reward Function ($\mathcal{R}$): The DRL agent is rewarded based on the improvement in aggregate network performance metrics relative to a fixed-parameter game formulation $(w_{\gamma 0}, w_{p 0})$. The reward $r_{(t)}$ at time step $t$ is given by:
    \begin{equation} \label{eq:reward}
    r_{(t)} = \Delta\bar\Gamma_{(t)} - \beta \cdot \Delta\bar P_{(t)}, 
    \end{equation}
    where $\Delta\bar\Gamma_{(t)}$ denotes the gain in aggregate normalized network SINR, and $\Delta\bar P_{(t)}$ denotes the change in aggregate normalized network power consumption at time step $t$. Coefficient $\beta$ is a positive weighting factor that determines the importance of reducing power consumption. Both $\Delta\bar\Gamma_{(t)}$ and $\Delta\bar P_{(t)}$ are calculated based on the resulting associations after each distributed game reaches equilibrium. Specifically:
    \begin{align}
    \Delta\bar\Gamma_{(t)} &= \sum_{u \in \mathcal{U}} \bar\Gamma^k_{u,b}(s^*_{u}(t)) - \sum_{u \in \mathcal{U}} \bar\Gamma^k_{u,b}(s^*_{u,0}(t)), \\
    \Delta\bar P_{(t)} &= \sum_{u \in \mathcal{U}} \bar {P}_b(s^*_{u}(t)) - \sum_{u \in \mathcal{U}} \bar {P}_b(s^*_{u,0}(t)).
    \end{align}
    Here, $s^*_{u}(t)$ represents the converged strategy of user $u$ under the agent's policy at time step $t$, and $s^*_{u,0}(t)$ the one under the fixed-parameter configuration $(w_{\gamma 0}, w_{p 0})$. 
\end{itemize}

Crucially, the proposed MDP formulation keeps the dimensionality of both the state and action spaces constant regardless of the number of users in the network, thereby enabling scalability. This contrasts with centralized RL and BS-agent MARL algorithms employed in UARA frameworks, which may define combinatorial action spaces, directly mapping UEs to BSs and radio resources, and consequently experience exponential action-space growth and high state-space dimensionality as network density increases. Moreover, the state space does not include user-specific information, such as converged associations, CSI data, complete SINR vectors, user coordinates, or user priority class IDs. Instead, the agent operates on BS load distribution data, thus limiting the exposure of UE-specific information to the controller.

\begin{algorithm}[t]
\caption{DQN-Based Network Orchestration Scheme for Game-Theoretic UARA}
\label{alg:dqn_metacontrol}

\textbf{Input:} Dataset of HetNet topologies, Action Space grid $\mathcal{A}$, Number of episodes $E$, Number of time steps $T$, Hyperparameters $(\alpha, \delta, \epsilon, \epsilon_{\min}, \epsilon_{\text{decay}})$, Mini-batch size $B$

\textbf{Output:} Learned DQN model $Q(x, a; \theta)$

\vspace{1mm}
\hrule
\vspace{1mm}

\textbf{Initialize:} Replay Buffer $\mathcal{D} \leftarrow \emptyset$, network weights $\theta$ randomly, exploration rate $\epsilon \leftarrow 1.0$

\For{each $episode = 1 \dots E$}{
    Load a network realization from dataset\;
    Select initial random action $a_{(0)}=(w_{\gamma(0)},w_{p(0)})$\;
    Execute Algorithm~\ref{alg:uara} to observe initial state $x_{(1)}$\;
    \For{each step $t = 1 \dots T$}{
        With probability $\epsilon$ select random $a_{(t)} \in \mathcal{A}$, else $a_{(t)} = \arg\max_a Q(x_{(t)},a;\theta)$\;
        Broadcast $a_{(t)}$ to all UEs\;
        Execute Algorithm~\ref{alg:uara} until PSNE is reached\;
        Receive reward $r_{(t)}$ using Eq. \eqref{eq:reward} and observe next state $x_{(t+1)}$\;
        Store transition $(x_{(t)}, a_{(t)}, r_{(t)}, x_{(t+1)})$ into $\mathcal{D}$; overwrite oldest if buffer is full\;
        \If{size of $\mathcal{D}$ $>$ $2B$}{
            Sample a random mini-batch of $B$ transitions $(x_{(j)}, a_{(j)}, r_{(j)}, x_{(j+1)})$ from $\mathcal{D}$\;   
            Calculate target values $y_j$ using Eq. \eqref{eq:target_val}\;   
            Perform a gradient descent step on $(y_j - Q(x_{(j)}, a_{(j)}; \theta))^2$:
            \[
            \theta \leftarrow \theta - \alpha \nabla_\theta \frac{1}{B}\sum_j (y_j - Q(x_{(j)}, a_{(j)}; \theta))^2
            \]
        }
        $x_{(t)} \leftarrow x_{(t+1)}$\;
    }
    Decay exploration rate: $\epsilon \leftarrow \max(\epsilon_{\min}, \epsilon \cdot \epsilon_{\text{decay}})$\;
}
\end{algorithm}

\subsection{DQN-Based Network Orchestration Scheme} 
Exploiting the discrete action space, we solve the formulated MDP using a DQN as a simple and efficient action-value approximator, thereby avoiding the architectural and computational complexity of policy-gradient methods. 
Specifically, the neural network with weights $\theta$ is trained to approximate $Q(x, a; \theta) \approx Q^*(x, a)$ by minimizing the Mean Squared Error (MSE) loss derived from the Bellman equation:
\begin{equation} \label{eq:loss}
    L(\theta) = \mathbb{E} \left[ \left( y_t - Q(x_{(t)}, a_{(t)}; \theta) \right)^2 \right],
\end{equation}
where $y_t$ is the target value defined as:
\begin{equation}  \label{eq:target_val}
    y_t = r_{(t)} + \delta \max_{a'} Q(x_{(t+1)}, a'; \theta),
\end{equation}
and $\delta \in [0,1]$ is the discount factor of the MDP representing the importance of future rewards. To improve DQN learning stability, we implement Experience Replay (ER). Specifically, transitions of the form $(x_{(t)}, a_{(t)}, r_{(t)}, x_{(t+1)})$ are stored in a replay buffer $\mathcal{D}$. During training, random mini-batches are sampled from $\mathcal{D}$ to approximate the expected MSE loss and update the weights of the NN via gradient descent. 

\begin{figure}[t]
\centering
\includegraphics[width=0.97\linewidth]{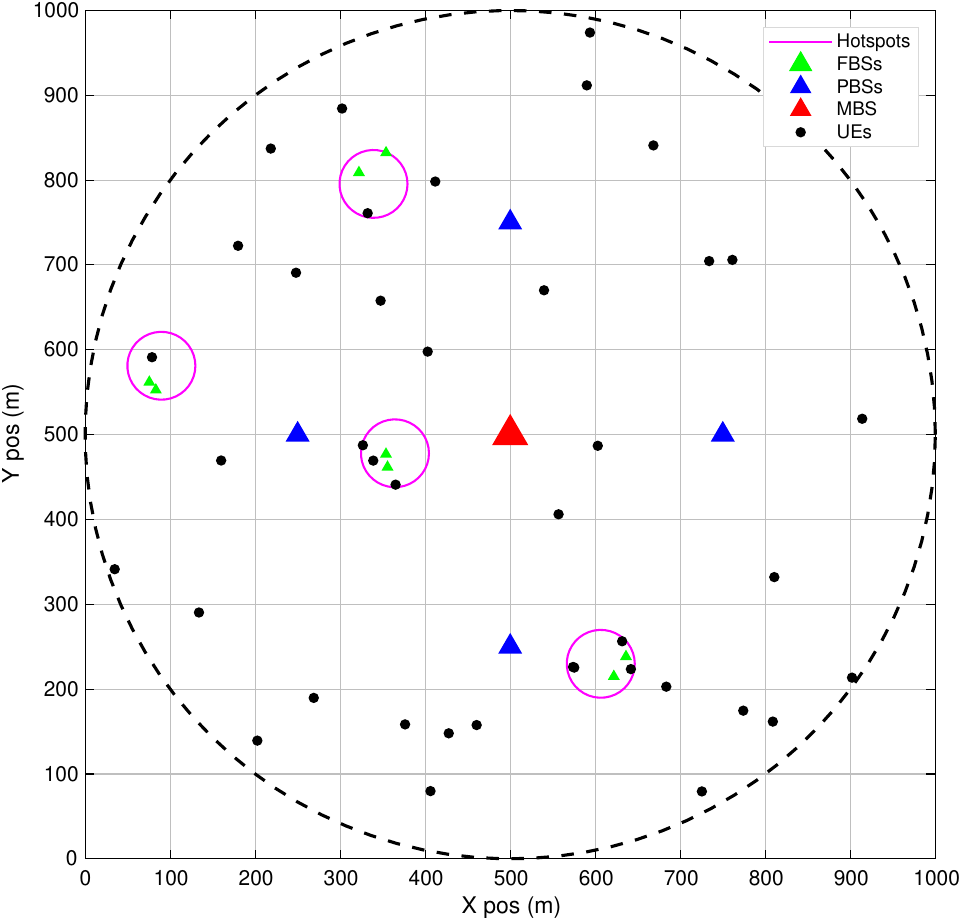}
\caption{Example topology of the simulated HetNet.}
\label{fig:hetnet_layout}
\end{figure}

The proposed DQN-based network orchestration
scheme, illustrated in Fig.~\ref{fig:arch}, is detailed in Alg.~\ref{alg:dqn_metacontrol}. First, the NN weights are randomly initialized, the replay buffer is set to empty, and the exploration rate is set to 1.0. At the beginning of each episode, a new network realization is loaded from the dataset. The agent selects a random initial action $a_{(0)}=(w_{\gamma(0)},w_{p(0)})$, and the UARA game is executed to obtain the initial state $x_{(1)}$. Each episode consists of $T$ steps. At each step $t$, the agent observes the state $x_{(t)}$ and selects, with probability $\epsilon$, a random action $a_{(t)} \in \mathcal{A}$, or else picks the action maximizing the approximated action-value 
function, i.e., $a_{(t)} = \arg\max_a Q(x_{(t)}, a; \theta)$. The selected action, i.e., the parameter configuration, is broadcast to all UEs and the game is played until convergence. Upon convergence, the agent receives its reward $r_{(t)}$ and observes the new state $x_{(t+1)}$. The transition $(x_{(t)}, a_{(t)}, r_{(t)}, x_{(t+1)})$ is stored in the replay buffer $\mathcal{D}$, overwriting the oldest entry if the buffer is full. Once the buffer exceeds $2B$ transitions (ensuring sufficient diversity), a random mini-batch of size $B$ is sampled and target values $y_j$ are computed using Eq. \eqref{eq:target_val}. A gradient descent step then minimizes Eq. \eqref{eq:loss} over the sampled mini-batch to update $\theta$. At the end of each episode, the exploration rate is decayed as $\epsilon \leftarrow \max(\epsilon_{\min}, \epsilon \cdot \epsilon_{\text{decay}})$. After $E$ episodes, the agent converges to a learned action-value function $Q(x, a; \theta)$ that approximates the optimal policy for the selection of the game-theoretic utility parameters.

Because each of the $T$ steps across the $E$ episodes involves a mini-batch gradient update, the DQN training complexity is $\mathcal{O}\big(E T B (N_{in} N_h + N_h N_{out})\big)$, where $N_{in}, N_h, \text{and } N_{out}$ denote the dimensions of the input, hidden, and output layers, respectively.
Conversely, during online deployment, the trained agent executes a single forward pass with complexity $\mathcal{O}(N_{in} N_h + N_h N_{out})$. Since the NN dimensions are independent of the number of users, the inference complexity remains $\mathcal{O}(1)$ with respect to $N$.

\begin{table}[t]
\captionsetup{justification=centering, labelsep=newline}
\caption{HetNet Simulation Parameters}
\label{tab:params}
\centering
\begin{tabular}{lc}
\toprule
\textit{Parameter} & \textit{Value} \\
\midrule
    \multicolumn{2}{l}{\textbf{General System Parameters}} \\[1pt]
Downlink center frequency $f$ & 2 GHz \\
Channel bandwidth $W$ & 180 kHz \\
Number of channels $K$ & 2 \\
Noise power spectral density $N_0$ & -174 dBm/Hz \\
Network area radius & 500 m \\
Hotspot zone radius & 40 m \\
Load fraction standard deviation $\sigma$ & 0.025 \\[3pt]
\multicolumn{2}{l}{\textbf{Base Stations}} \\[1pt]
Number of PBSs, FBSs & 4, 8 \\
Transmit power (MBS, PBSs, FBSs) & \{40, 30, 20\} dBm \\
Antenna heights (MBS, PBSs, FBSs) & \{25, 10, 5\} m \\
BS antenna model & 3GPP-3D \\
Propagation scenario (MBS) & 3GPP 38.901 UMa \\
Propagation scenario (PBSs/FBSs) & 3GPP 38.901 UMi \\[3pt]
\multicolumn{2}{l}{\textbf{User Equipment}} \\[1pt]
Number of UEs $N$ & 40 \\
Minimum hotspot UE ratio & 20\% \\
Priority UE ratio & 20\% \\
Priority weights $\omega_u$ (standard, priority) & \{1, 4\} \\
UE antenna height & 1.5 m \\
UE antenna model & Patch \\
\bottomrule
\end{tabular}
\end{table}

\begin{table}[t]
\captionsetup{justification=centering, labelsep=newline}
\caption{DQN Hyperparameters}
\label{tab:hyperparams}
\centering
\begin{tabular}{lc}
\toprule
\textit{Parameter} & \textit{Value} \\
\midrule
Episodes $E$ & 500 \\
Steps $T$ & 30 \\
Replay Memory $D$ Size & 3000 \\
Mini-batch size $B$ & 32 \\
Learning rate $\alpha$ & 0.001 \\
Discount factor $\delta$ & 0.8 \\
Initial exploration rate $\epsilon$ & 1.0 \\
Minimum exploration rate $\epsilon_{\min}$ & 0.05 \\
Exploration decay $\epsilon_{\text{decay}}$ & 0.992 \\
\bottomrule
\end{tabular}
\end{table}

\section{Performance Evaluation}\label{sec:per_eval}
A dataset of 500 HetNet instances is produced for training, each containing the full SINR matrix 
$\gamma_{u,b}^k, \; \forall (u, b, k) \in \mathcal{U} \times 
\mathcal{B} \times \mathcal{K}$. These channel realizations are generated using the QuaDRiGa simulator\cite{quadriga}, following standardized 3GPP TR 38.901 propagation models\cite{3gpp38901} and capturing path loss, shadowing, and fast fading effects. An example network topology of the dataset is illustrated in Fig.~\ref{fig:hetnet_layout}. All dataset instances use the same BS layout and environment seed, ensuring spatially consistent large-scale propagation effects across the simulations. Meanwhile, $N$ UEs are placed randomly in each instance to model the spatial variability of traffic distribution in HetNets. UEs are also distributed with a higher concentration within Hotspot Zones. These simulate high traffic areas of practical HetNets (e.g., public event areas) and are  primarily served by the FBSs. Furthermore, UEs are divided into two priority classes—standard and priority. The network simulation parameters are explicitly defined in Table~\ref{tab:params}. The tier-dependent power cost $\bar{P}_b$ in Eq. \eqref{eq:util} is obtained by normalizing the transmit power of each BS tier in decibels against the power of the Macro-tier. This results in fixed power costs of 0.5, 0.75, and 1.0 for the Femto, Pico, and Macro tiers, respectively.

\begin{figure}[t]
\centering
\includegraphics[width=\linewidth]{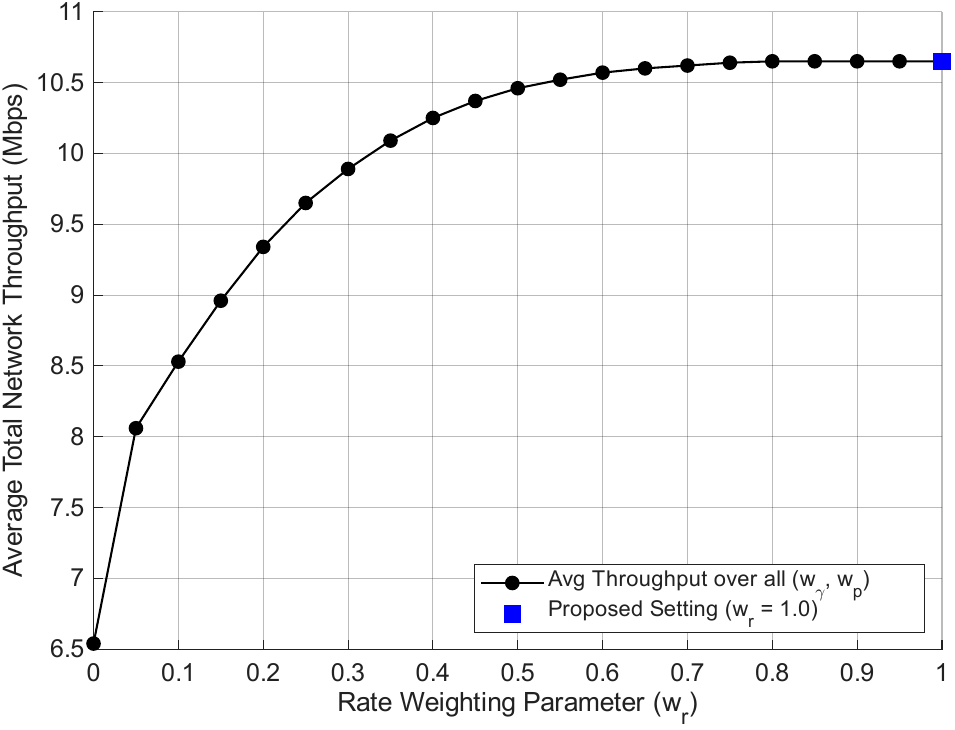}
\caption{Impact of the rate weighting parameter $w_r$ on the average total network throughput.}
\label{fig:wr_throughput}
\end{figure}

\begin{figure}[t]
\centering
\includegraphics[width=\linewidth]{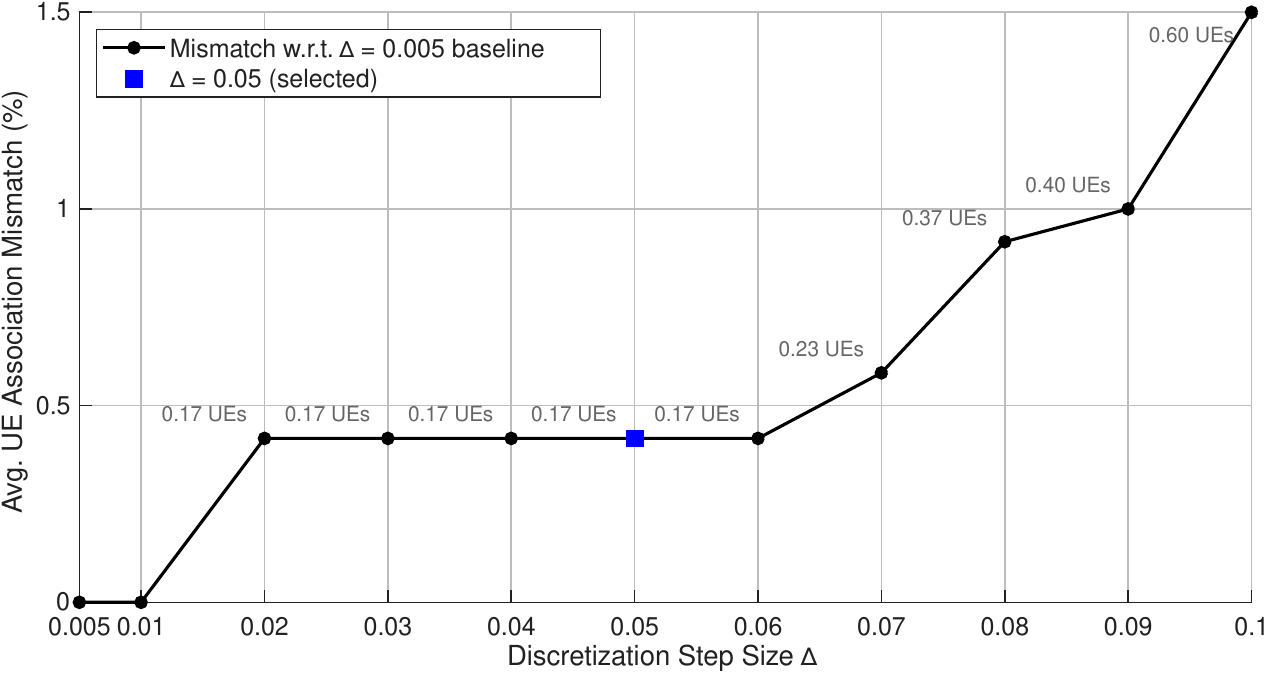}
\caption{Impact of the action space discretization step $\Delta$ on user association accuracy.}
\label{fig:resolution}
\end{figure}

The proposed DQN consists of an input layer of $N_{in} = 15$ neurons (13 BS load values and 2 utility parameters), a single hidden layer of $N_h = 256$ neurons with ReLU activation, and an output layer of $N_{out} = |\mathcal{A}|$ neurons. Training is performed using mini-batch stochastic gradient descent. The complete set of DQN hyperparameters is provided in Table~\ref{tab:hyperparams}.

\subsection{Simulation Results} \label{subsec:results}
This subsection evaluates the proposed framework under different operational objectives. First, the rate weighting parameter $w_r$ of the UARA game is calibrated. Then, the action-space discretization and training convergence of the DQN agent are examined. Finally, the proposed DQN-orchestrated game-theoretic framework (DQN-GT) is evaluated against conventional association methods, a static game formulation, and an exhaustive-search benchmark under balanced, power-aware, and coverage-enhancing policies.

\subsubsection{Rate Weighting Parameter Calibration}
As discussed in Section~\ref{sec:mdp}, the centralized agent configures $(w_{\gamma}, w_{p}) \in [0, 1]^2$, while the value of $w_{r}$ is fixed. Fig.~\ref{fig:wr_throughput} illustrates how the rate weighting parameter $w_r$ of the game's utility function affects total network throughput. For each value of $w_r \in [0, 1]$, the distributed UARA game is executed across all configurations
of $(w_{\gamma},w_p)$ in the discrete grid and the average total network throughput is calculated. As shown in the figure, this metric is maximized for $w_r \geq 0.8$. Therefore, fixing $w_r$ to 1.0 forces UEs to prioritize connections that provide the
highest possible data rates under each $(w_\gamma,w_p)$ configuration. In this way, the controller can implement different operational policies while maintaining robust user throughputs, which is a primary objective in cellular networks. Accordingly, $w_r$ is fixed to $1.0$ for the remaining simulations in this work.

\begin{figure}[t]
\centering
\includegraphics[width=\linewidth]{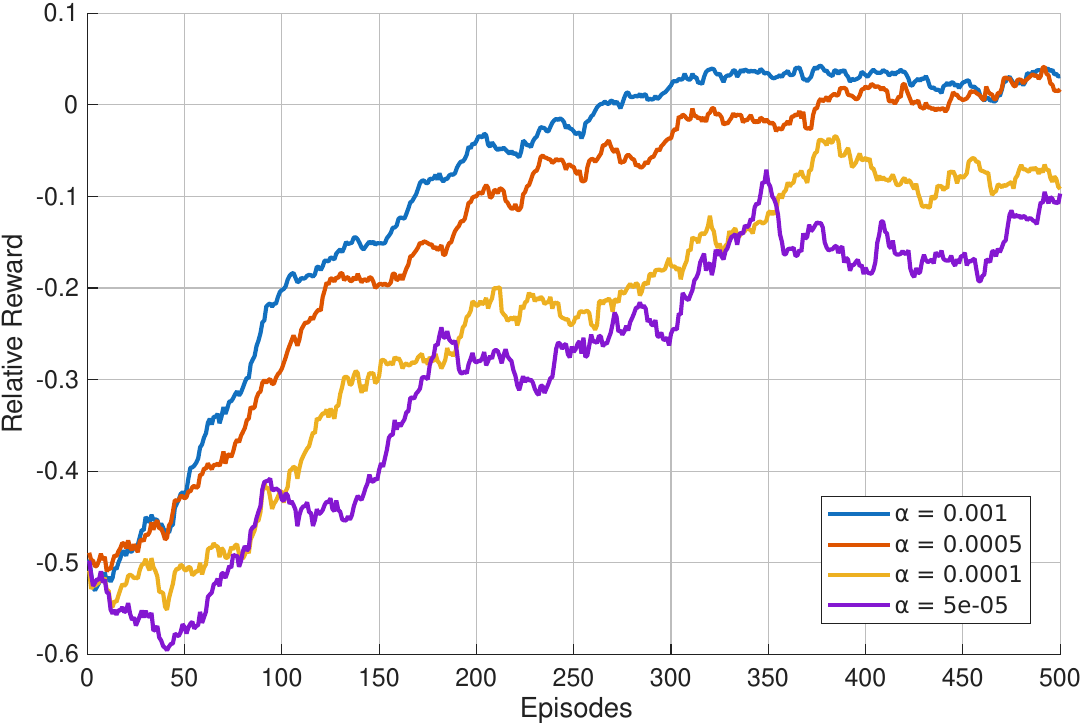} 
\caption{Training convergence of the proposed DQN algorithm under varying learning rates $\alpha$ (50-episode moving average of the relative reward).}
\label{fig:reward}
\end{figure}

\subsubsection{Action Space Discretization and Convergence Behavior}
The DQN agent is initially trained for the balanced policy, which enforces an effective trade-off between network coverage and power consumption. The fixed-parameter baseline and the reward coefficient are set to $(w_{\gamma 0}, w_{p 0}) = (0.5, 0.3)$ and $\beta = 0.2$, respectively.

The action space discretization step $\Delta$ in Eq. \eqref{eq:step} must first be selected to balance solution accuracy and computational complexity. Fig.~\ref{fig:resolution} illustrates the average number of UEs that change their converged association relative to the high-resolution baseline ($\Delta = 0.005$), with respect to the grid step size. The evaluation is conducted using exhaustive search.  Step sizes up to $\Delta = 0.06$ introduce a negligible deviation of 0.17 UEs on average (0.43\%), while larger steps lead to a rapid increase in association mismatch. Therefore, $\Delta=0.05$ is selected as it provides near-baseline accuracy and uniformly partitions $[0,1]$, while yielding an action space of $|\mathcal{A}|=441$ configurations. Hence, the output layer of the proposed DQN consists of 441 neurons.

Fig.~\ref{fig:reward} evaluates the training convergence of the proposed DQN-based network orchestration algorithm under different learning rates $\alpha$, using a 50-episode moving average. Training under $\alpha = 0.001$ provides the highest final reward  ($\approx 0.035$) and converges to a stable policy by episode 300, while training under lower rates ($\alpha \in \{5 \times 10^{-4}, 1 \times 10^{-4}, 5 \times 10^{-5}\}$) converges more slowly or fails to reach competitive performance. Therefore, $\alpha = 0.001$ is selected for the proposed scheme.

\begin{table*}[t]
\centering
\footnotesize
\captionsetup{justification=centering, labelsep=newline}
\caption{Average System Performance Metrics Relative to the Balanced Policy}
\label{tab:balance}
\begin{tabular}{l S S S S S S S S S S S S}
\toprule
\multirow{2}{*}{Method} & \multicolumn{4}{c}{$N = 20$} & \multicolumn{4}{c}{$N = 40$} & \multicolumn{4}{c}{$N = 80$} \\
\cmidrule(lr){2-5} \cmidrule(lr){6-9} \cmidrule(lr){10-13}
 & {\makecell{SINR\\(dB)}} & {\makecell{Power\\(W)}} & {\makecell{Throughput\\(Mbps)}} & {\makecell{Response\\(ms)}} & {\makecell{SINR\\(dB)}} & {\makecell{Power\\(W)}} & {\makecell{Throughput\\(Mbps)}} & {\makecell{Response\\(ms)}} & {\makecell{SINR\\(dB)}} & {\makecell{Power\\(W)}} & {\makecell{Throughput\\(Mbps)}} & {\makecell{Response\\(ms)}} \\
\midrule
Max-SINR & 7.35 & 61.02 & 6.07 & {\raisebox{0.15ex}{<}} 0.01 & 7.40 & 117.15 & 8.98 & {\raisebox{0.15ex}{<}} 0.01 & 7.30 & 239.16 & 11.04 & {\raisebox{0.15ex}{<}} 0.01 \\
CRE & 6.97 & 47.76 & 6.29 & {\raisebox{0.15ex}{<}} 0.01 & 7.01 & 88.11 & 9.09 & {\raisebox{0.15ex}{<}} 0.01 & 6.90 & 178.61 & 10.85 & {\raisebox{0.15ex}{<}} 0.01 \\
ES-GT & 6.95 & 48.40 & 7.31 & 49.73 & 6.89 & 87.22 & 10.24 & 89.79 & 6.74 & 173.59 & 11.19 & 186.05 \\
DQN-GT & 6.98 & 50.76 & 7.26 & 0.17 & 6.87 & 90.42 & 10.20 & 0.27 & 6.52 & 178.14 & 11.22 & 0.48 \\
Static Game & 6.55 & 44.34 & 7.46 & 0.14 & 5.78 & 68.66 & 10.37 & 0.24 & 4.87 & 120.94 & 10.94 & 0.45 \\
\bottomrule
\end{tabular}
\end{table*}

\begin{table*}[t]
\centering
\footnotesize 
\captionsetup{justification=centering, labelsep=newline}
\caption{Average System Performance Metrics of the Power-Aware Policy}
\label{tab:power}
\begin{tabular}{l S S S S S S S S}
\toprule
\multirow{2}{*}{Method} & \multicolumn{4}{c}{$N = 20$} & \multicolumn{4}{c}{$N = 40$} \\
\cmidrule(lr){2-5} \cmidrule(lr){6-9} 
& {\makecell{SINR\\(dB)}} & {\makecell{Power\\(W)}} & {\makecell{Throughput\\(Mbps)}} & {\makecell{Response\\(ms)}}& {\makecell{SINR\\(dB)}} & {\makecell{Power\\(W)}} & {\makecell{Throughput\\(Mbps)}} & {\makecell{Response\\(ms)}} \\
\midrule
ES-GT & 2.10 & 23.19 & 7.27 & 47.64 & 1.32 & 34.44 & 10.83 & 89.41 \\
DQN-GT & 3.27 & 27.86 & 7.57 & 0.18 & 1.51 & 36.83 & 10.94 & 0.30 \\
\bottomrule
\end{tabular}
\end{table*}

\begin{table*}[t]
\centering
\footnotesize 
\captionsetup{justification=centering, labelsep=newline}
\caption{Average System Performance Metrics of the Coverage-Enhancing Policy}
\label{tab:coverage}
\begin{tabular}{l S S S S S S S S}
\toprule
\multirow{2}{*}{Method} & \multicolumn{4}{c}{$N = 40$} & \multicolumn{4}{c}{$N = 80$} \\
\cmidrule(lr){2-5} \cmidrule(lr){6-9} 
& {\makecell{SINR\\(dB)}} & {\makecell{Power\\(W)}} & {\makecell{Throughput\\(Mbps)}} & {\makecell{Response\\(ms)}}& {\makecell{SINR\\(dB)}} & {\makecell{Power\\(W)}} & {\makecell{Throughput\\(Mbps)}} & {\makecell{Response\\(ms)}} \\
\midrule
ES-GT & 7.29 & 104.46 & 10.14 & 88.90 & 7.22 & 217.85 & 11.52 & 187.92 \\
DQN-GT & 7.19 & 102.55 & 10.20 & 0.30 & 7.14 & 213.98 & 11.45 & 0.49 \\
\bottomrule
\end{tabular}
\end{table*}

\subsubsection{Comparative Performance Evaluation}
Finally, the performance of the proposed DQN-GT is evaluated against four baseline methods:
\begin{itemize}
    \item Max-SINR association: UEs select the connection providing the highest SINR.
    
    \item Cell Range Expansion (CRE): fixed artificial biases are applied to the SINR received from small cells ($+6$ dB to PBSs and $+11$ dB to FBSs).
    
    \item Exhaustive search-orchestrated game (ES-GT): the UARA game is simulated for all candidate utility parameter configurations to identify the optimal one.
    
    \item Static Game Formulation: utility parameters are fixed to baseline values, $(w_{\gamma 0}, w_{p 0}) = (0.5, 0.3)$.
\end{itemize}
Performance is evaluated with respect to average UE SINR, total network power consumption, aggregate network throughput, and response time of each method. The simulations are conducted on a Windows 10 Pro system with an Intel Core i5 CPU and 16 GB of memory.

Table~\ref{tab:balance} presents the comparative results under the balanced operational policy of the controller. The tier-specific means are set to $\mu_b = \{0.125, 0.075, 0.075\}$ for the MBS, PBSs, and FBSs, respectively. Max-SINR inherently achieves the highest average UE SINR across all network sizes. However, this comes at the cost of overloading higher-power BSs, resulting in the highest network power consumption and consistently low network throughput across all evaluated traffic densities. The controller-orchestrated game-theoretic frameworks achieve performance close to CRE in terms of average UE SINR and network power consumption, while delivering higher network throughput. Specifically, relative to CRE in the 20-UE case, ES-GT records a 0.02 dB decrease in average UE SINR and a 1.34\% increase in power consumption, while achieving 16.22\% higher  throughput than CRE and 20.43\% higher than Max-SINR. This demonstrates the advantage of the proposed multi-objective game formulation over conventional signal-based association methods in terms of resource management. Similar results are observed across the 40- and 80-UE cases, with throughput gains gradually decreasing, however, as the network becomes more congested and optimization margins narrow.

The proposed DQN-GT closely approximates the performance of ES-GT in the communication metrics evaluated for $N=40$, demonstrating high accuracy in the implementation of the optimal policy. A high level of accuracy is also maintained for traffic loads of 20 and 80 users, which were not encountered during training. This stems from the agent operating on BS load distributions, which capture macroscopic network performance patterns and therefore enable operation across different traffic loads. Such a property contrasts with many RL models employed in HetNet UARA schemes, which are designed for specific user counts and typically require retraining to adapt to different loads.

Crucially, the proposed DQN-GT provides a significant advantage over ES-GT in terms of execution speed. The total DQN-GT response time accounts for both the DQN inference time and the convergence time of the underlying UARA game. In the 20-UE case, DQN-GT achieves a response time approximately 293 times shorter than ES-GT. This gap widens with network density, since the convergence time of the UARA game grows with the number of players, and ES-GT must simulate it separately for every candidate parameter configuration. Specifically, in the 80-UE case, DQN-GT operates approximately 388 times faster than ES-GT. 

Nonetheless, even in the case of the highly congested network, where 80 UEs contend for 26 available BS-channel pairs, the proposed DQN-GT records a response time of 0.48 ms. For reference, the channel coherence time of a vehicular 
receiver with a speed of 45 km/h, operating at a 2 GHz downlink frequency, is approximately 5.08 ms, confirming that the proposed framework operates well within real-time constraints in terms of algorithmic execution. This is a direct effect of its bilevel architecture and the lightweight structure of the DQN. Specifically, the proposed DQN maintains an $\mathcal{O}(1)$ inference complexity with respect to network load, yielding a rapid inference time of 0.03 ms regardless of traffic density. This is obtained from Table~\ref{tab:balance} as the difference between DQN-GT and Static Game response times. While these execution times are highly promising, it should be noted that they reflect the total algorithmic complexity of the proposed framework and do not account for practical deployment overhead. In real-world systems, some additional latency would arise from signaling overhead required to aggregate BS load observations and broadcast the updated utility parameters, as well as from UEs transmitting local SINR values to receive back global extremes for the normalization of the utility components.

In contrast to centralized RL and BS-agent MARL algorithms, which can suffer from exponential growth of the action space as the number of users increases, the proposed framework decouples the dynamic policy update from the distributed UARA execution. By offloading the combinatorial associations problem to a fast-converging Nash game with linear per-iteration complexity, DQN-GT incurs low computational overhead even in dense traffic conditions, enabling scalability. Moreover, the underlying UARA game requires only simple algebraic calculations from UEs to evaluate each link. This eliminates the need for executing locally deep NNs, as in MARL frameworks that may model UEs as independent agents, thereby enabling UARA execution even on computationally and energy-constrained devices (e.g., IoT). 

In the 20-UE network, the Static Game achieves results close to both CRE and the controller-orchestrated game-theoretic frameworks in terms of average UE SINR and network power consumption. However, as network density increases, a progressively larger  deviation in these metrics is observed.  Specifically, while the Static Game incurs a small gap of 0.42 dB in average UE SINR relative to CRE for $N = 20$, this gap exceeds 2 dB for $N = 80$. Notably, signal-based association methods remain inherently robust to congestion: users associate with the BS-channel pair providing the highest SINR (or highest biased SINR) regardless of network load. Consequently, this progressive deviation relative to CRE highlights the limitation of a static game formulation in providing consistent performance across different congestion levels. In contrast, the controller-orchestrated frameworks consistently approximate CRE in terms of coverage and power consumption across the evaluated traffic densities, demonstrating comparable robustness to network congestion.  

The power-aware policy is trained by setting $(w_{\gamma 0}, w_{p 0}) = (0.5, 0.5)$, $\beta = 0.8$, and $\mu_b = \{0.075, 0.075, 0.125\}$.  Table~\ref{tab:power} presents the performance of DQN-GT and ES-GT under this operational objective. For $N = 40$, ES-GT effectively implements power-aware operation. Specifically, total network power consumption is reduced by $60.51\%$ relative to the Balanced Policy results of Table~\ref{tab:balance}, and by $70.60\%$ relative to Max-SINR. By offloading users toward small cells in order to save power, the congestion in the macro-tier BSs is also reduced. Consequently, total network throughput is increased relative to the balanced operation, as well, by $5.76\%$. In the 20-UE case, power consumption is again effectively reduced; however, no gain in network throughput is observed. In this low-congestion scenario, the abundance of available resources (20 users contending for 26 BS-channel pairs) means that UEs can establish lightly loaded connections largely independent of their converged strategy. As expected, improvements in network power consumption and throughput come at the cost of a lower, although still viable, average UE SINR compared to the Balanced Policy.

In the 40-UE network the proposed DQN-GT closely approximates the exhaustive-search baseline in the evaluated metrics of average UE SINR, total network power consumption, and aggregate throughput. In the case of the previously unseen load of $N = 20$ UEs, it also effectively implements the power-aware policy, exhibiting, however, a deviation of 1.17 dB in average UE SINR and 4.67 Watts in total power consumption relative to ES-GT. This slight trade-off in accuracy is the cost of deploying DQN-GT to unseen traffic loads and of its significant response time advantage over ES-GT, which is consistent with the results reported under the Balanced Policy. Specifically, similarly to the Balanced Policy, under the Power-Aware Policy DQN-GT records $0.18$ ms for $N = 20$ and $0.30$ ms for $N = 40$, while ES-GT requires $47.64$ ms and $89.41$ ms, respectively, corresponding to a speedup of approximately $265\times$ and $298\times$ in favor of the proposed framework.

The proposed Power-Aware Policy is particularly well-suited for low-congestion scenarios. For instance, during periods of reduced traffic, such as late-night hours, the network controller can shift the objective to power-aware operation. By increasing the penalty on high-power connections, the proposed framework reduces power consumption while consistently providing viable connections and high data rates for UEs in scenarios where channel resources are abundant, such as the $N = 20$ case examined above.

To train the DQN agent for the Coverage-Enhancing Policy, we set $(w_{\gamma 0}, w_{p 0}) = (0.5, 0.2)$, $\beta = 0.06$, and $\mu_b = \{0.20, 0.10, 0.05\}$. The corresponding performance metrics of ES-GT and DQN-GT are presented in Table~\ref{tab:coverage}. Under this operational objective, ES-GT effectively enhances coverage relative to the Balanced Policy. Specifically, for both $N = 40$ and $N = 80$, it provides higher average UE SINR than CRE, by 0.28 dB and 0.32 dB, respectively, approaching the performance of Max-SINR. It also achieves higher aggregate throughput than both signal-based methods across the evaluated densities, and than the Balanced Policy for the highly congested $N=80$ case. Meanwhile, it reduces power consumption by $10.83\%$ for $N=40$ and by $8.91\%$ for $N=80$ relative to Max-SINR. 

The proposed DQN-GT achieves high accuracy in the implementation of the optimal policy, approximating the exhaustive search baseline in terms of average UE SINR, total network power consumption, and aggregate network throughput, consistent with the behavior observed under the other operational policies. This holds for both the trained user density of $N = 40$ and the previously unseen density of $N = 80$, further indicating the adaptability of the proposed framework. Meanwhile, DQN-GT achieves a significantly faster response than ES-GT, providing a relative speedup of approximately $296\times$ and $384\times$ for $N = 40$ and $N = 80$, respectively.

The proposed Coverage-Enhancing Policy is particularly well-suited for high-congestion scenarios. For instance, during periods of sudden traffic surges, such as the conclusion of public events, the controller can shift the operational objective to coverage enhancement. By prioritizing signal quality over power savings, DQN-GT provides reliable connections and higher data rates for UEs in scenarios where radio resources are limited, such as the $N = 80$ case examined above.

\section{Conclusion}\label{sec:conclusions}
In this paper, we presented a novel DQN-based scheme for the orchestration of game-theoretic User Association and Resource Allocation in HetNets. The proposed low-complexity DQN controller dynamically configures the utility parameters of the distributed game, effectively transitioning the network between power awareness, coverage enhancement, and balanced operation in direct response to macroscopic traffic patterns. Simulations in a practical HetNet environment demonstrated that, under the considered operational objectives, the proposed bilevel framework efficiently manages network resources across different traffic densities, while incurring low computational overhead. As a future direction, the proposed framework could be extended to RIS-assisted HetNets, examining the integration of programmable surfaces for communication enhancement. 

\bibliographystyle{ieeetr}
\bibliography{refs}

\end{document}